\documentclass[pra, 12 pt]{revtex4}
\usepackage[english]{babel}
\usepackage{amsmath}
\usepackage{epsfig}

\begin{document}

\title{\bf GROUND STATE ENERGY IN THE EXTERNAL FIELD AND THE PROBLEM OF DENSITY FUNCTIONAL APPROXIMATIONS
}
\author{V.B. Bobrov $^{1}$, S.A. Trigger $^{1,2}$}
\address{$^1$ Joint\, Institute\, for\, High\, Temperatures, Russian\, Academy\,
of\, Sciences, Izhorskaia St., 13, Bd. 2. Moscow\, 125412, Russia;\\
emails:\, vic5907@mail.ru,\;satron@mail.ru\\
$^2$ Eindhoven  University of Technology, P.O. Box 513, MB 5600
Eindhoven, The Netherlands}

\begin{abstract}
Based on the Schrodinger equation, exact expressions for the non-relativistic particle energy in the local external field and the external field potential are derived as inhomogeneous density functionals. On this basis, it is shown that, when considering more than two noninteracting electrons, the energy of such a system cannot be an inhomogeneous density functional. The result is extended for the system of interacting electrons. This means that the Hohenberg-Kohn lemma which assert that in the ground state to each inhomogeneous density corresponds only one potential of the external field cannot be a justification of the existence of the universal density functional in the general case. At the same time, statements of the density functional theory remain valid when considering any number of noninteracting ground-state bosons due to the Bose condensation effect.\\

 PACS number(s): 31.15.E-, 71.15.Mb, 52.25.-b, 05.30.Fk\\

\end{abstract}

\maketitle
According to the Hohenberg--Kohn lemma [1,2], the same inhomogeneous density $ n({\bf r})$ cannot correspond to two different local potentials $v_1({\bf r})$ and $v_2({\bf r})$ of the external field in the ground state of the non-relativistic system of electrons (except the case $ v_1({\bf r}) - v_2({\bf r}) = \mathrm{const} $). Thus, the inhomogeneous density $n({\bf r})$ of the ground-state non-relativistic electron system uniquely defines the potential $ v({\bf r})$ (to within an additive constant). In the case of ground state degeneration, the lemma relates to the density $ n ({\bf r})$ of any ground state.
It is generally accepted [1] that from the Hohenberg--Kohn lemma the external field potential $ v({\bf r})$ is an inhomogeneous density functional,
\begin{eqnarray}
n({\bf r})=n[v({\bf r})] \to v({\bf r})= v[n({\bf r})] + \mathrm{const}.
 \label{F1}
\end{eqnarray}
Existence of the functional (the square brackets in (1)) means that there is a universal rule according to which the external field potential $ v({\bf r})$ can be found by the known inhomogeneous density $ n({\bf r})$ corresponding to the ground state of a system under consideration. This means that in principle there exists (although cannot be explicitly found or indicated) the rule of determining the function $ v({\bf r})$ by the known function $ n({\bf r})$ whose structure is independent of the explicit form of functions $ v({\bf r})$ and $ n({\bf r})$. We emphasize, that this very strong statement has not yet been called into question. In fact, in the general case, there is no one-to-one correspondence between the Hohenberg--Kohn lemma and relation (1) on functional dependence. It only follows from the Hohenberg--Kohn lemma that quite a definite external field $ v({\bf r})$ can be put in correspondence with each function $ n({\bf r})$ (within a constant factor). However, this does not mean that such a correspondence is established by the unified rule $ v({\bf r})=v[n({\bf r})]$ universal for any external field [3].

In other words, each external field determines the unique density (it is clear, e.g., on the basis of the universal rules of
the perturbation theory), and each density determines the unique external field on the basis of the Hohenberg--Kohn lemma. However, the rule for the last correspondence can be non-universal. This rule in general depends on the concrete form of the density.
Opportunity of such non-universality of the rule of correspondence for the "inverse" functional $ v({\bf r})=v[n({\bf r})]$ was not considered in [1].  Existence of this non-universality violates the Hohenberg-Kohn theorem, although the Hohenberg-Kohn lemma is undoubtedly correct. Below we show that the universal density functional does not exist in the case of the number of fermions more than two.

For this reason, let us pay attention to the fact that the inhomogeneous density $ n({\bf r})$ is the functional $n[v({\bf r})]$ by definition $ n({\bf r})=\langle\Psi_0 \mid \Psi^+({\bf r})\,\Psi({\bf r})\mid \Psi_0\rangle$ and the functional dependence of the wave function $\Psi_0$ of the ground state on the external field $v({\bf r})$, $\Psi_0=\Psi_0 [v({\bf r})]$. Here $\Psi^+({\bf r})$ and $\Psi({\bf r})$ are the field creation and annihilation operators. Thus, the correspondence rule between the density $ n({\bf r})$ and field $ v({\bf r})$ is established based on the solution of the corresponding Schrodinger equation for the wave function $ \Psi_0$ in a given external field
$v({\bf r})$ and the inhomogeneous density $ n({\bf r})$ determination. Then, it only follows from the lemma proved by Hohenberg and Kohn [1] that the functional $n[v({\bf r})]$ is unique (taking into account the condition $n[v({\bf r})]=n[v({\bf r})+const]$). Here it is clear that the functional $n[v({\bf r})]$ is essentially nonlinear in the external field $v({\bf r})$. This means that two possibilities are admissible without violating the Hohenberg--Kohn lemma: (i) the inverse problem on the determination of the dependence of $v({\bf r})$ on $n({\bf r})$ has individual solutions to each pair of functions $n({\bf r})$ and $v({\bf r})$ (or for certain types (classes) of pairs of functions $n({\bf r})$ and $v({\bf r})$); (ii) the inverse problem has the universal solution $ v({\bf r}) = v[n({\bf r})]$. As noted above, this dilemma is not usually considered, and it is assumed that there is the universal solution $ v[n({\bf r})]$ valid for any external field and any number of particles, i.e., the possibility (ii) is always realized [3].

In fact, as follows from the above, the Hohenberg--Kohn lemma is insufficient for the statement about the existence of the universal solution $v[n ({\bf r})]$. However, it seems impossible to disprove the statement about the universality in the general form. In this regard, we use the "proof to the contrary". Let us assume that the functional $ v({\bf r}) = v[n ({\bf r})]$ exists and analyze consequences of this statement. By the example of noninteracting fermions, let us show that such an assumption leads to contradiction.

If we accept the validity of statement (1), then the energy $E_0$ the ground-state system of $N$ interacting electrons with the Hamiltonian $H$ in the external field with potential $ v({\bf r})$, which is characterized by the wave function $\Psi_{0}$, can be written as
\begin{eqnarray}
E_0\equiv \langle\Psi_{0}|H|\Psi_{0}\rangle = E_0(N,[ \Psi_{0},v({\bf r})])=E_0[n({\bf r}),v({\bf r})],\qquad N=\int n({\bf r})d^3{\bf r}.
\label{F2}
\end{eqnarray}

Here it is taken into account that $\langle\Psi_{0}|\Psi_{0} \rangle =1$ and $\Psi_{0}[v({\bf r})] = \Psi_{0}[v({\bf r}) + \mathrm{const}]$. In turn, it immediately follows from (2) that the quantity $F[n({\bf r})] = \langle\Psi_{0} |T+U|\Psi_{0} \rangle $ which defines the system ground state energy
\begin{eqnarray}
E_0[n({\bf r}),v({\bf r})]=F[n({\bf r})]+\int v({\bf r})n({\bf r})d^3{\bf r}
, \label{F3}
\end{eqnarray}
is the functional of only the density $n({\bf r})$ ("universal" density functional). In this case, it is accepted to use the term "universal" in the sense of the independence of the explicit form of the external field potential [1,2], although this necessitates to consider the $v$-representability of the inhomogeneous density.

Here the operators $T$ and $U$ are the operators of the kinetic and interparticle interaction energies, respectively. Statement (3) is the basis of the density functional theory (DFT) widely used in various areas of physics and chemistry (see, e.g., [4,5]). However, the exact form of this universal functional is still unknown even for noninteracting electrons $(U=0)$. It is clear that there is one-to-one correspondence between the statements about the existence of the functionals $v[n({\bf r})]$ and $F [n({\bf r})]$, i.e., the existence of one of the these functionals predetermines the existence of another. Hence, if the functional $F[n ({\bf r})]$ in (3) does not exist, the functional $v[n ({\bf r})]$ also does not exist [3].

In this connection, let us consider one non-relativistic electron of mass $m$ in the static external field $v({\bf r})$. Then the electron steady state characterized by a certain set of quantum numbers $\alpha $, including the spin quantum number $\sigma $, is completely defined by wave function $\Phi_{\alpha}({\bf r})$ which satisfies the Schrodinger equation
\begin{eqnarray}
\left\{-\frac{\hbar ^2}{2m}\Delta_{{\bf r}}+v({\bf r}) \right\}\Phi_{\alpha}({\bf r})= \epsilon_{\alpha} \Psi_{\alpha }({\bf r}),
\label{F4}
\end{eqnarray}
where $\epsilon_{\alpha}$ is the electron energy in a corresponding state. Since the electron energy is independent of spin, each value of $\epsilon_{\alpha}$ is doubly degenerate in the spin quantum number, as well as other values of physical quantities, including the inhomogeneous density $n_{\alpha}({\bf r})$. It is accepted to solve equation (4) for eigenvalues when the boundary condition $\Phi_{\alpha}(|{\bf r}|\to \infty)=0$ (the so-called condition at infinity [6]) is satisfied. Taking into account the possibility of considering the system in a finite volume $V$, the boundary condition for Eq. (4) in the most general form is written as

\begin{eqnarray}
\Phi_{\alpha}({\bf r}\to {\bf S})=0,
 \label{F5}
\end{eqnarray}
where $S$ is the surface bounding the volume $V$. We further take into account that the wave function $\Phi_{\alpha}({\bf r})$ can be considered as a real function [6]. Then
\begin{eqnarray}
n_{\alpha}({\bf r})=| \Phi_{\alpha}({\bf r})|^2= \Phi_{\alpha}^2({\bf r}),
\label{F6}
\end{eqnarray}

\begin{eqnarray}
\nabla_{\bf r}n_{\alpha}({\bf r})=2\Phi_{\alpha}({\bf r})\nabla_{\bf r}\Phi_{\alpha}({\bf r}),.\qquad \Delta_{\bf r}n_{\alpha}({\bf r})=2\Phi_{\alpha}({\bf r})\Delta_{\bf r}\Phi_{\alpha}({\bf r})+2(\nabla_{\bf r}\Phi_{\alpha}({\bf r}))(\nabla_{\bf r}\Phi_{\alpha}({\bf r})).
\label{F7}
\end{eqnarray}
It immediately follows from (4)-(7) that the inhomogeneous density $n_{\alpha}({\bf r}) $ satisfies the equation for eigenvalues

\begin{eqnarray}
-\frac{\hbar ^2}{4m}\Delta_{\bf r}n_{\alpha}({\bf r})+\frac{\hbar ^2}{8mn_{\alpha}({\bf r})} (\nabla_{\bf r}n_{\alpha}({\bf r}))(\nabla_{\bf r}n_{\alpha}({\bf r}))+v({\bf r})n_{\alpha}({\bf r})= \epsilon_{\alpha}n_{\alpha}({\bf r})
\label{F8}
\end{eqnarray}
with the boundary conditions
\begin{eqnarray}
n_{\alpha}({\bf r\to S})=0, \qquad \nabla_{\bf r}n_{\alpha}({\bf r})|_{\bf r\to S}=0.
\label{F9}
\end{eqnarray}
Now let us integrate Eq. (8) over the volume occupied by the system, taking into account the normalization condition immediately following from (6),
\begin{eqnarray}
\int n_{\alpha}({\bf r})dV=1
\label{F10}
\end{eqnarray}
Then, from (8), we find the density functional for the energy $\epsilon_{\alpha}$
\begin{eqnarray}
\epsilon_{\alpha}[n_{\alpha}({\bf r}),v({\bf r})]= F^{(1)}[n_{\alpha}({\bf r})]+\int v({\bf r}) n_{\alpha}({\bf r})dV
\label{F11} ,
\end{eqnarray}

\begin{eqnarray}
F^{(1)} [n_{\alpha}({\bf r})]= F^{(1)}_0 [n_{\alpha}({\bf r})]+ F^{(1)}_W[n_{\alpha}({\bf r})],
\label{F12}
\end{eqnarray}

\begin{eqnarray}
F^{(1)}_0[n_{\alpha}({\bf r})]= -\frac{\hbar ^2}{4m}\int \Delta_{\bf r}n_{\alpha}({\bf r})dV, \qquad F^{(1)}_W[n_{\alpha}({\bf r})]=\frac{\hbar ^2}{8m}\int \frac{(\nabla_{\bf r}n_{\alpha}({\bf r}))(\nabla_{\bf r}n_{\alpha}({\bf r}))}{n_{\alpha}({\bf r})} d V.
\label{F13}
\end{eqnarray}
Here $F^{(1)} [n_{\alpha}({\bf r})]$ is the universal density functional $F[n({\bf r})] $ (3) for one particle (index (1)). The functional $F^{(1)} [n_{\alpha}({\bf r})]$ is written in (12) in the form of two terms by two reasons. First, taking into account the Gauss formula and the second boundary condition in (9), the functional $F^{(1)}_0[n_{\alpha}({\bf r})]$ vanishes,
\begin{eqnarray}
F^{(1)}_0[n_{\alpha}({\bf r})]= -\frac{\hbar ^2}{4m}\int \Delta_{\bf r}n_{\alpha}({\bf r})dV= -\frac{\hbar ^2}{4m}\oint \nabla_{\bf r}n_{\alpha}({\bf r})d{\bf S}=0.
\label{F14}
\end{eqnarray}
Second, the functional explicit form $F^{(1)}_W [n_{\alpha}({\bf r})]$ (13) formally is exactly identical to the so-called Weizsaker correction to the Thomas--Fermi kinetic energy functional [7] (see [4,5] for more details). In contrast to the Weizsacker correction, the expression $F^{(1)}_W [n_{\alpha}({\bf r})]$ (13) in this problem is an exact expression for the \,"universal"\, functional of the inhomogeneous density for one particle.
Thus, in the case at hand, the \,"universal"\, density functional $F^{(1)} [n_{\alpha}({\bf r})]$ exists, is found exactly, and is written as
\begin{eqnarray}
F^{(1)}[n_{\alpha}({\bf r})]= F^{(1)}_W[n_{\alpha}({\bf r})]
\label{F15}
\end{eqnarray}
By direct calculation (see, e.g., [5]), it is easy to verify that
\begin{eqnarray}
\frac{\delta F^{(1)}_W[n_{\alpha}({\bf r})]}{\delta n_{\alpha}({\bf r})} = - \frac{\hbar ^2}{4mn_{\alpha}({\bf r})}\Delta_{\bf r}n_{\alpha}({\bf r})+ \frac{\hbar ^2}{8mn_{\alpha}^2 ({\bf r})} (\nabla_{\bf r}n_{\alpha}({\bf r})) (\nabla_{\bf r}n_{\alpha}({\bf r})).
\label{F16}
\end{eqnarray}
Thus, Eq. (8) for energy eigenvalues $\epsilon^{(1)}$ of one particle in the external field $v({\bf r})$ with boundary conditions (9) is a consequence of the variation equation for the energy $\epsilon^{(1)} [n^{(1)}({\bf r}),v({\bf r})] $ as the inhomogeneous density functional $n^{(1)}({\bf r})$ of one particle in the specified external field $v({\bf r})$,
\begin{eqnarray}
\delta \epsilon^{(1)}[n^{(1)}({\bf r})]=0
\label{F17}
\end{eqnarray}
Indeed, using normalization condition (10) and the Legendre transform, from (17), we find
\begin{eqnarray}
\frac{\delta \epsilon^{(1)} [n^{(1)}({\bf r})]}{\delta n^{(1)}({\bf r})} = \mathrm{const}.
\label{F18}
\end{eqnarray}
To determine the constant in Eq. (18), we take into account that, according to (11)-(16),
\begin{eqnarray}
\frac{\delta F^{(1)}[n^{(1)}({\bf r})]}{\delta n^{(1)}({\bf r})} + v({\bf r}) = \mathrm{const} = \epsilon^{(1)}.
\label{F19}
\end{eqnarray}
Thus, the variation equation (17) is equivalent to (8) and is completely identical to the corresponding equation of the particle system as the wave function functional in quantum mechanics (see, e.g., [6]). Hence, in the case of one particle under consideration, there exists the density functional $v[n ({\bf r})]$ for the external field potential, determined within a constant factor,
\begin{eqnarray}
v [n^{(1)}({\bf r})]+ \mathrm{const}=\frac{\delta F^{(1)}[n^{(1)}({\bf r})]}{\delta n^{(1)}({\bf r})} = \frac{\hbar ^2}{4m n^{(1)}({\bf r})} \Delta_{\bf r}n^{(1)}({\bf r})-\frac{\hbar ^2}{8m[n^{(1)}({\bf r})]^2}(\nabla_{\bf r}n^{(1)}({\bf r})) (\nabla_{\bf r}n^{(1)}({\bf r})).
\label{F20}
\end{eqnarray}
Thus, as noted above, the existence of the functional $F^{(1)} [n^{(1)}({\bf r})]$ (15) predetermines the existence of the functional $v [n({\bf r})]$ (20), and vice versa.

Taking into account that the Hohenberg--Kohn lemma proof, as well as statements (1) and (3), are by no means independent of a particular value of the number of particles $N$ in the system under study, the assumption on the existence of the \,"universal"\, density functional makes it possible to extend the results obtained to the case of an arbitrary number of noninteracting electrons.
Then the explicit form of the functional for any number of particles is obtained using the following replacements:
\begin{eqnarray}
n^{(1)}({\bf r})\to n({\bf r}),\qquad \epsilon^{(1)}\to E_0.
\label{F21}
\end{eqnarray}
In this case, the normalization condition (10) is replaced by the corresponding condition in (2), and its is also accepted that, according to (3), (11)-(13), that the quantity $\langle\Psi_{0}|T|\Psi_{0} \rangle $ for the system of several noninteracting particles is written as
\begin{eqnarray}
\langle\Psi_{0}|T|\Psi_{0} \rangle = F^{(0)}[n({\bf r})]=-\frac{\hbar ^2}{4m}\int \Delta_{\bf r}n({\bf r})dV +\frac{\hbar ^2}{8m}\int \frac{(\nabla_{\bf r}n({\bf r}))(\nabla_{\bf r}n({\bf r})) }{n({\bf r})} dV.
\label{F22}
\end{eqnarray}
To ascertain, are relations (21) and (22) valid, we consider a system of $N$ noninteracting electrons in the external field $v({\bf r})$.
To account for the identity of electrons, it is the most convenient to perform such a consideration in the secondary quantization formalism (see [4-6] for more details). We note that such consideration is fully equivalent to the use of
Slater determinants to describe the wave function of the system of noninteracting electrons and to implement the Young scheme [5,6]. Then, any state of the system of $N$ noninteracting identical electrons is characterized by a set of the so-called "occupied" single-particle states $\alpha_1, \ldots, \alpha_N$ (see (4)) and, by virtue of the Pauli principle,
\begin{eqnarray}
\alpha_i \neq   \alpha_j   \quad \textrm{ for }\quad i\neq j.
\label{F23}
\end{eqnarray}
Then the energy $E^{(0)}$ and inhomogeneous density $n^{(0)}({\bf r})$ in a corresponding state are given by [5,6]
\begin{eqnarray}
E^{(0)}(\alpha_1, \ldots, \alpha_N) = \sum_{\alpha _i}\epsilon_{\alpha_i}, \qquad n^{(0)}({\bf r}, \alpha_1, \ldots, \alpha_N)= \sum_{\alpha _i}n_{\alpha_i}({\bf r}).
\label{F24}
\end{eqnarray}
In this case, $ \epsilon_{\alpha_i} = \epsilon_{\alpha_i} [n_{\alpha_i}({\bf r}), v({\bf r})]$ (see (11)-(13)), and $ \Sigma_{\alpha_i}1=N$.
Then, it immediately follows from (23) and (24) that the energy $E^{(0)} $ of the system of $N(N\geq 3)$ noninteracting identical electrons, including the ground state energy $E^{(0)}_0$, cannot be the density functional $n^{(0)}({\bf r}) $ in the specified external field due to the nonlinearity of the functional $F^{(1)}_W[n_{\alpha_i} ({\bf r})] $ (13). It is clear that a similar statement also takes place for the universal functional $\langle\Psi_{0}|T|\Psi_{0} \rangle $. This is a consequence of the fact that the external field potential $v({\bf r})$ at $N\geq 3$ cannot be presented as the density functional $n({\bf r})$ (see (20)).

In the case of two noninteracting ground-state electrons, the DFT statements remain valid due to the double degeneracy in the spin quantum number ($n({\bf r}) = 2n^{(1)}({\bf r})$, see, for example, [8]). We also note that these results are directly associated with the Pauli principle which is related only to fermions. In the case of ground-state bosons which are "accumulated" at one lowest energy level (Bose condensation), the DFT statements remain valid at an arbitrary number of noninteracting bosons.

Thus, for more than two noninteracting fermions in the inhomogeneous ground state, the external field potential $v({\bf r})$ is not the density functional $n({\bf r})$, i.e., $v({\bf r})\neq v[n ({\bf r})]$.

Let us show now that the obtained results lead straightforward to validation of the analogous conclusion for the system of interacted electrons. The Hamiltonian of the system of inhomogeneous electrons can be written as
\begin{eqnarray}
H_\lambda=T+\int d {\bf r} v({\bf r}) \hat{n}({\bf r}) +\lambda U.
\label{F25}
\end{eqnarray}
where $\hat{n}({\bf r})$  is the operator of the electron density, $\langle\Psi_{0}\mid\hat{n}({\bf r})\mid\Psi_{0} \rangle=n({\bf r})$ and $U$ is the operator of the electron interaction energy. Then the energy $E_0(\lambda)$  of the ground state $\Psi_{0}(\lambda)$  satisfies to the equality (see, e.g., [9])
\begin{eqnarray}
\frac{\partial E_0(\lambda)}{\partial\lambda}=\langle\Psi_{0}(\lambda)\mid \frac{\partial H_{\lambda}}{\partial\lambda}\mid\Psi_{0}(\lambda)\rangle=\langle\Psi_{0}(\lambda)\mid U \mid\Psi_{0}(\lambda)\rangle.
\label{F26}
\end{eqnarray}
From (26) straightforward follows the explicit relation for the ground state energy of the system of interacting electrons
\begin{eqnarray}
E_0-E_0^{(0)}=\int_0^1 \frac{d \lambda}{\lambda}\langle U \rangle_\lambda.
\label{F27}
\end{eqnarray}
where $E_0^{(0)}$  is the ground state energy of the inhomogeneous electron gas without interaction, which has been calculated above, $\langle U \rangle_\lambda$  is the average potential energy of inhomogeneous electron system with the Hamiltonian (25). Take further into account that if the density functional $v[n(r)]$  exists, the average kinetic energy, as well as average potential energy, of the inhomogeneous electron system is the universal density functional (singly) [10]. It is obvious, that presence of the parameter  $\lambda$  in the Hamiltonian (25) has no any influence on this statement as well as the integration in Eq. (27). Then from (27) directly follows that for the existence of the universal density functional $F[n({\bf r})]$   (see (3)) it is necessary the existence of the universal density functional for the inhomogeneous electron system without interaction. This is impossible, as it was shown above, for more then two Fermi particles. In a different way the impossibility of existence of the density functional of inhomogeneous electron system follows from the fact that the terms in the row of the perturbation theory for energy, which contain interaction potential, have another nature than the terms without interaction. Therefore, they cannot compensate the "non-universality" of the kinetic energy of non-interacting electrons.

As a result, we come to the conclusion that the Hohenberg--Kohn lemma [1,2] cannot be a justification of the existence of the "universal" density functional as an exact statement or a theorem. At the same time, in any approximations (e.g., in the limit of weak inhomogeneity of the external field or in the quasi-classical limit for the electron gas), the \,"universal"\, functional can exist.
In this connection, let us pay attention to the following circumstance. The main approximation for  density functional construction is the so-called Local Density Approximation (LDA) (for detail, see [4,5,10]). The basis of LDA is the dependence of the energy of homogeneous electron gas on average density $n$ which equals to the ratio of the full number $N$ of electrons to the volume $V$, $n=N/V$. Consideration of the homogeneous electron gas, it one's turn, based on use of the thermodynamic limit transition $N\rightarrow\infty, V\rightarrow\infty, N/V\rightarrow n\neq 0$ (see, e.g., [9]). This means that the model of homogeneous electron system cannot be used as the initial approach for consideration of a finite quantity of electrons in an external filed (in particular, in the case of electrons in the field of one or several nuclei, when the conditions of the thermodynamic limit transition are not valid even for a large value $Z\gg 1$, where $Z$ is the nuclear charge).

\section*{Acknowledgment}

This study was supported by the Netherlands Organization for Scientific Research (NWO), project no. 047.017.2006.007 and the Russian Foundation for Basic Research, projects no. 07-02-01464-a and no. 10-02-90418-Ukr-a.

\end{document}